\documentclass[useAMS,usenatbib]{mn2e}
\usepackage{graphicx}

\title[The complex light-curve of the afterglow of GRB\,071010A]{The complex light-curve of the afterglow of GRB\,071010A\thanks{Based on observations made also with ESO Telescopes at the La Silla and Paranal Observatory under programmes IDs 080.D-0229, 080.D-0250 and 080.D-792.}
}

\author[S. Covino et al.]{S. Covino,$^1$
\thanks{E-mail: stefano.covino@brera.inaf.it} 
P. D'Avanzo,$^{1,2}$ A. Klotz,$^{3,4}$  D.A. Perley,$^5$ L. Amati,$^6$ S. Campana,$^1$  \newauthor G. Chincarini,$^{7,1}$ A. Cucchiara,$^8$ V. D'Elia,$^{9}$ D. Guetta,$^{9}$ C. Guidorzi,$^{1,7}$  D.A. Kann,$^{10}$ \newauthor A. K\"upc\"u Yolda\c{s},$^{11}$ K. Misra,$^{12,13}$ G. Olofsson,$^{14}$  G. Tagliaferri,$^1$ L.A. Antonelli,$^{9}$ \newauthor E. Berger,$^{15}$ J.S. Bloom,$^5$ M. B\"oer,$^3$ C. Clemens,$^{11}$ F. D'Alessio,$^9$ M. Della Valle,$^{16,17,18}$ \newauthor S. di Serego Alighieri,$^{19}$ A.V. Filippenko,$^5$ R.J. Foley,$^5$ D.B. Fox,$^7$D. Fugazza,$^1$  \newauthor J. Fynbo,$^{20}$ B. Gendre,$^{21}$ P. Goldoni,$^{22,23}$ J. Greiner,$^{11}$  D. Kocevksi,$^5$ E. Maiorano,$^6$ \newauthor N. Masetti,$^6$  E. Meurs,$^{24,25}$  M. Modjaz,$^5$ E. Molinari,$^1$ A. Moretti,$^1$   E. Palazzi,$^6$ \newauthor S.B. Pandey,$^{12}$  S. Piranomonte,$^{9}$ D. Poznanski,$^5$ N. Primak,$^{11}$ P. Romano,$^1$ E. Rossi,$^{26}$ \newauthor R. Roy,$^{12}$  J.M. Silverman,$^5$ L. Stella,$^{9}$ G. Stratta,$^{27}$ V. Testa,$^{9}$ S.D. Vergani,$^{24,25}$ \newauthor F. Vitali,$^{9}$ F. Zerbi,$^1$\\
$^1$ INAF/Osservatorio Astronomico di Brera, via Bianchi 46, 23807, Merate (LC), Italy\\
$^2$ Universit\`a dell'Insubria, Dipartimento di Fisica e Matematica, via Valleggio 11, 22100 Como, Italy\\
$^3$ Observatoire de Haute-Provence, 04870 Saint-Michel l'Observatoire, France\\
$^4$ CESR, 9 Avenue colonel Roche, Universit\'e de Toulouse, 31400 Toulouse, France\\
$^5$ Astronomy Department, University of California, 445 Campbell Hall, Berkeley, CA 94720-3411, USA\\
$^6$ INAF/Istituto di Astrofisica Spaziale e Fisica Cosmica di Bologna, via Gobetti 101, 40129 Bologna, Italy\\
$^7$ Universit\`a degli Studi di Milano, Bicocca, Piazza delle Scienze 3, 20126, Milano, Italy\\
$^8$ Department of Astronomy and Astrophysics, Pennsylvania State University, USA\\
$^{9}$ INAF/Osservatorio Astronomico di Roma, via Frascati 33, 00040 Monteporzio Catone (Roma), Italy\\
$^{10}$ Th\"uringer Landessternwarte Tautenburg, Sternwarte 5, 07778 Tautenburg, Germany\\
$^{11}$ Max-Planck-Institut f\"ur extraterrestrische Physik, Giessenbachstrasse 1, 85748 Garching, Germany\\
$^{12}$ Aryabhatta Research Institute of observational sciences (ARIES), Manora Peak, Nainital 263 129, India\\
$^{13}$ Inter University Center for Astronomy and Astrophysics, Post Bag 4, Ganeshkhind, Pune 411 007 India\\
$^{14}$ Stockholm Observatory, Roslagstullsbacken 21, 10691 Stockholm, Sweden\\
$^{15}$ Observatories of the Carnegie Institution of Washington, 813 Santa Barbara Street, Pasadena, CA 91101, USA\\
$^{16}$ INAF/Osservatorio Astronomico di Capodimonte, Via Moiariello 16, 80131 Napoli, Italy\\
$^{17}$ Icranet, International Center for Relativistic Astrophysics Network, Piazza Repubblica 10, Pescara, Italy \\
$^{18}$ European Southern Observatory, Garching bei Munchen -Germany\\
$^{19}$ INAF/Osservatorio Astrofisica di Arcetri, Largo Enrico Fermi 5, 50125 Firenze, Italy\\
$^{20}$ Dark Cosmology Centre, Niels Bohr Institute, University of Copenhagen, Juliane Maries vej 30, 2100 Kobenhavn, Denmark\\
$^{21}$ Laboratoire d'Astrophysique de Marseille/CNRS/Universit\'e de Provence, 13376 Marseille Cedex 12, France\\
$^{22}$ Laboratoire Astroparticule et Cosmologie, 10 rue A. Domon et L. Duquet, 75205 Paris Cedex 13, France\\
$^{23}$ Service d'Astrophysique, DSM/DAPNIA/SAp, CEA-Saclay, 91191 Gif-sur-Yvette, France\\
$^{24}$ Dunsink Observatory - DIAS, 31 Fitzwilliam Street, Dublin 2, Ireland\\
$^{25}$ School of Physical Sciences and NCPST, Dublin City University, Dublin 9, Ireland\\
$^{26}$ JILA, University of Colorado, 440 UCB Boulder, CO 80309-0440, USA\\
$^{27}$ ASI Science Data Center (ASDC), via G. Galilei, 00044 Frascati, Italy\\
}

\begin{document}

\date{}

\pagerange{\pageref{firstpage}--\pageref{lastpage}} \pubyear{2008}

\maketitle

\label{firstpage}

\begin{abstract}
We present and discuss the results of an extensive observational
campaign devoted to GRB\,071010A, a long-duration gamma-ray burst
detected by the \textit{Swift} satellite. This event was followed for
almost a month in the optical/near-infrared (NIR) with various
telescopes starting from about 2\,min after the high-energy
event. \textit{Swift}-XRT observations started only later at about
0.4\,d. The light-curve evolution allows us to single out an
initial rising phase with a maximum at about 7\,min, possibly the
afterglow onset in the context of the standard fireball model, which
is then followed by a smooth decay interrupted by a sharp
rebrightening at about 0.6\,d. The rebrightening was visible in both
the optical/NIR and X-rays and can be interpreted as an episode of
discrete energy injection, although various alternatives are possible. 
A steepening of the afterglow light curve
is recorded at about 1\,d. The entire evolution of the
optical/NIR afterglow is consistent with being achromatic. This could
be one of the few identified GRB afterglows with an achromatic break
in the X-ray through the optical/NIR bands. Polarimetry was also
obtained at about 1\,d, just after the rebrightening and almost
coincident with the steepening. This provided a fairly tight upper
limit of 0.9\% for the polarized-flux fraction.
\end{abstract}


\section{Introduction}

The {\it Swift} satellite \citep{Geh04} has revolutionised gamma-ray
burst (GRB) science by introducing dedicated observational strategies
and providing a wealth of high-quality data. In particular, a major
breakthrough has been the spectral and temporal coverage at X-ray
energies from a few minutes up to weeks or months after the GRB. This
has provided unprecedented information on the late prompt and early
afterglow evolution \citep{Nou06}. Optical observations from the
ground, however, have not always been of comparable quality and
late-time phases have been seldom monitored accurately
\citep{Cov06}. This is a consequence of the strong increase in the
number of events to be followed, the large average redshift of
\textit{Swift} GRBs \citep{Berg06,Jak06}, coupled with the
difficulties in performing productive follow-up observations from the
ground.  While the early-time afterglow is often well sampled by the
network of small and medium-class robotic telescopes around the world,
the late-time afterglow can only be monitored with the largest
instruments and only at the cost of significant observing time
\citep{Dai07}. Based on these considerations, we started a new program
of afterglow observations with the explicit goal of providing full
coverage of the optical/near-infrared (NIR) afterglow light curve for
a selected number of events (building on visibility and availability
of early-time observations) rather than providing sparse datapoints
for most of the observable {\it Swift} GRBs.

The importance of dense temporal and spectral sampling is particularly
crucial for the identification and interpretation of the different
features in the light-curve evolution that are observed both in the
optical \citep{Kann07} and at higher energies \citep{ZhMe04}. In
particular, in spite of intense observational efforts and continuous
theoretical interest
\citep[e.g.,]{BuRa06,Cov06,Cur07,Dad07,Dai07,Pan08}, the search for
light-curve jet breaks fully satisfying the requirements of the
standard fireball model, and thus allowing a correct inference of the
true energy content of a GRB, is still far from complete
\citep{Lian07}.

The GRB\,071010A afterglow evolution was followed in the optical by
TAROT and in the NIR by REM and GROND at early times, then by
Gemini-N, GROND, NOT, TNG, NTT, Keck-I, Sampurnanand, and VLT in the
subsequent two weeks. Multi-band observations were acquired to study
the temporal evolution of the afterglow spectral energy distribution
(SED). A polarimetric observation was carried out almost in
coincidence with a slope transition in the light curve. {\it Swift}
XRT and BAT data were also studied. In Sect.\,\ref{sec:grb} we report
information about GRB\,071010A, in Sect.\,\ref{sec:dana} our
observations, reductions, and data analysis are described, and in
Sect.\,\ref{sec:disc} a full discussion is presented. Our main
conclusions are summarised in Sect.\,\ref{sec:end}.

\section{GRB\,071010A}
\label{sec:grb}

GRB\,071010A was detected by {\it Swift} on Oct. 10, 2007 at 03:41:12
\citep{Mor07} (UT dates are used throughout this paper). The BAT light
curve was relatively broad, with a $T_{90}$ duration of $6 \pm 1$\,s
(a moderate-duration long GRB). The fluence was $(2.0 \pm 0.4) \times
10^{-7}$\,erg\,cm$^{-2}$ \citep[uncertainty 90\% confidence,
][]{Kri07}. Automatic slewing to the GRB position with the
\textit{Swift} narrow-field instruments was disabled, so the XRT
observations were delayed by 34\,ks \citep{Gui07b}. The optical
afterglow was promptly identified by TAROT \citep[$\alpha_{\rm J2000}$
= 19:12:14.624, $\delta_{\rm J2000}$ = $-$32:24:07.16; ][]{Blo07},
revealing an initial flux increase up to about 470\,s after the
high-energy event, followed by a regular decay
\citep{Klo07a,Klo07b}. Spectroscopy was carried out with Keck
\citep{Pro07}, revealing a strong Mg\,II absorber and corresponding
Fe\,II lines at redshift $z \approx 0.98$. A refined analysis was
carried out by \citet{Del08}, deriving $z = 0.985 \pm 0.005$
($1\sigma$ error) for the absorbing system with the highest
redshift. Lines of Mg\,II (2796--2803~\AA), Mg\,I (2852~\AA), Si\,II
(2336~\AA), Fe\,II (2367, 2374, 2586, 2600~\AA), and Mn\,II (2594,
2606~\AA) were identified. The afterglow was not detected in the radio
band at 8.46\,GHz almost two days after the GRB
\citep{ChFr07}. Late-time observations led \citet{Per07} to suggest
that the afterglow decay underwent a (possibly achromatic) steepening
about one day after the burst. Galactic dust absorption along the GRB
line of sight is $E_{\rm B-V} \approx 0.098$\,mag \citep{Sch98}.

Throughout this paper, the time decay and energy spectral indices
$\alpha$ and $\beta$ are defined by $F(t,\nu) \propto
(t-t_0)^{-\alpha}\nu^{-\beta}$, where $t_0$ is the trigger time of the
burst. We assume a $\Lambda$CDM cosmology with $\Omega_{\rm m} =
0.27$, $\Omega_\Lambda = 0.73$, and $h_0 = 0.71$. At the redshift of
the GRB the luminosity distance is $6.6$\,Gpc ($\sim 2.0 \times
10^{28}$\,cm), corresponding to a distance modulus $\mu =
44.1$\,mag. All uncertainties are $1\sigma$ unless stated otherwise.

\section{Observations, reductions, and analysis}
\label{sec:dana}

\subsection{Observations and reduction}

The afterglow of GRB\,071010A was observed by the TAROT telescope
\citep{Bri99} with the white and red filters, by the REM telescope
\citep{Zer01} equipped with the REMIR NIR camera \citep{Vit03,Con04},
and by the 2.2~m MPI/ESO telescope equipped with GROND \citep{Grei08}
starting from 124\,s after the GRB (86\,s after reception of the GCN
alert). The observations began at an initial airmass of about 2.3 with
the target already setting. In order to obtain an acceptable
signal-to-noise ratio (S/N) we needed to bin the REM data, thus
decreasing the highest time resolution afforded by the
instrument. Later NIR observations were carried out with Gemini-N
equipped with NIRI, the TNG equipped with NICS, and the NTT equipped
with SofI. In the optical the afterglow was again observed a few
hours after the GRB with the Keck-I telescope equipped with LRIS, the
Sampurnanand telescope, the NOT equipped with PolCor, and the VLT with
FORS1 and FORS2. Polarimetry was carried out with the VLT equipped
with FORS1 on Oct. 11, 2007 at 1:11:33 (0.89607\,d after the GRB)
with a total exposure time of 1800\,s. 

Data reduction was carried out following standard recipes with the
{\it Eclipse} package \citep{Dev97}. Photometry was computed by
aperture and profile-fitting photometry with the {\it Daophot}
\citep{Ste87} and {\it SExtractor} \citep{BeAr95}
packages. Photometric calibration was derived from the observation of
standard stars in the optical on two different nights and with the
2MASS catalogue \citep{Stru06} in the NIR, although in a few cases the
small field of view of NIR detectors made the calibration more
difficult. TAROT data were calibrated with observations of two field
stars and final magnitudes are expressed in the R2 system of the
USNO-B1 catalogue \citep{Mon03}. The results from the photometric
observations are reported in Table\,\ref{tab:photo}.

For polarimetry, the offset between instrumental and intrinsic
polarization was fixed by the observation of a polarized standard
star. Instrumental polarization was derived by observation of
non-polarized stars. The procedure we followed for the analysis of the
FORS1 polarization data was extensively discussed by
\citet{Cov99,Cov02,Cov03}. At the time of the observation the
afterglow optical emission was consistent with null polarization with
a 95\% upper limit of 0.9\% (1.3\% at 3\,$\sigma$) for the degree of polarized flux.

\begin{table*}
\begin{tiny}
\centering
\caption{Optical/NIR observations of GRB\,071010A. The reference time $t_{\rm GRB}$ is Oct. 10, 2007 at 03:41:12 \citep{Mor07}. Data are not corrected for dust absorption. For the
TAROT data, the ``clear'' magnitudes are calibrated against the $R$
filter. Data are sorted according to wavelength and according to time
for each filter.}
\label{tab:photo}
\begin{tabular}{ccccccc}
\hline
Mean date         & $t-t_{\rm GRB}$ & Exp time  & Airmass & Filter & Instrument & Magnitude      \\
(UT)              & (day)          & (s)       &         &        &            &                \\ 
\hline
2007 Oct 11.03601 &  0.88240 & 60$\times$2     & 1.2     & $U$    & VLT+FORS2  & $20.77\pm0.07$ \\
2007 Oct 12.01151 &  1.85790 & 60$\times$1     & 1.1     & $U$    & VLT+FORS2  & $22.11\pm0.13$ \\ 
2007 Oct 11.03945 &  0.88584 & 60$\times$2     & 1.2     & $B$    & VLT+FORS2  & $21.16\pm0.04$ \\
2007 Oct 12.02197 &  1.86836 & 60$\times$5     & 1.1     & $B$    & VLT+FORS2  & $22.75\pm0.05$ \\
2007 Oct 10.84027 &  0.68666 & 0.1$\times$6000 & 3.9     & $V$    & NOT+PolCor & $20.20\pm0.09$ \\
2007 Oct 11.03376 &  0.88015 & 40$\times$1     & 1.2     & $V$    & VLT+FORS1  & $20.35\pm0.04$ \\
2007 Oct 11.03880 &  0.88519 & 225$\times$1    & 1.2     & $V$    & VLT+FORS1  & $20.34\pm0.04$ \\
2007 Oct 11.04164 &  0.88803 & 60$\times$1     & 1.2     & $V$    & VLT+FORS2  & $20.35\pm0.04$ \\
2007 Oct 11.04188 &  0.88827 & 225$\times$1    & 1.2     & $V$    & VLT+FORS1  & $20.36\pm0.04$ \\
2007 Oct 11.04496 &  0.89135 & 225$\times$1    & 1.2     & $V$    & VLT+FORS1  & $20.36\pm0.04$ \\
2007 Oct 11.04804 &  0.89443 & 225$\times$1    & 1.2     & $V$    & VLT+FORS1  & $20.37\pm0.04$ \\
2007 Oct 11.05135 &  0.89774 & 225$\times$1    & 1.2     & $V$    & VLT+FORS1  & $20.39\pm0.04$ \\
2007 Oct 11.05441 &  0.90080 & 225$\times$1    & 1.3     & $V$    & VLT+FORS1  & $20.40\pm0.04$ \\
2007 Oct 11.05748 &  0.90387 & 225$\times$1    & 1.3     & $V$    & VLT+FORS1  & $20.40\pm0.04$ \\
2007 Oct 11.06056 &  0.90695 & 225$\times$1    & 1.2     & $V$    & VLT+FORS1  & $20.42\pm0.04$ \\
2007 Oct 11.20925 &  1.05564 & 180$\times$1 & 1.7      & $V$    & Keck\_I+LRIS & $20.62\pm0.04$\\
2007 Oct 11.21238 &  1.05877 & 180$\times$1 & 1.7      & $V$    & Keck\_I+LRIS & $20.62\pm0.04$\\
2007 Oct 12.02682 &  1.87321 & 60$\times$2     & 1.1     & $V$    & VLT+FORS2  & $21.94\pm0.04$ \\
2007 Oct 13.00848 &  2.85487 & 60$\times$2     & 1.1     & $V$    & VLT+FORS2  & $22.75\pm0.04$ \\
2007 Oct 10.20107 &  0.04745 & 100$\times$1 & 1.7      & $R$    & Keck\_I+LRIS & $18.15\pm0.06$ \\
2007 Oct 10.20351 &  0.04990 & 100$\times$1 & 1.7      & $R$    & Keck\_I+LRIS & $18.25\pm0.04$ \\
2007 Oct 10.28570 &  0.13209 & 120$\times$1 & 2.2      & $R$    & Keck\_I+LRIS &  $18.94\pm0.04$ \\
2007 Oct 10.28884 &  0.13523 & 30$\times$1 & 2.3      & $R$    & Keck\_I+LRIS & $18.98\pm0.04$ \\
2007 Oct 10.29230 &  0.13869 & 30$\times$1 & 2.3      & $R$    & Keck\_I+LRIS & $18.97\pm0.04$ \\
2007 Oct 10.29375 &  0.14014 & 30$\times$1 & 2.4      & $R$    & Keck\_I+LRIS &  $18.98\pm0.04$ \\
2007 Oct 10.29530 &  0.14169 & 30$\times$1 & 2.4      & $R$    & Keck\_I+LRIS & $19.00\pm0.04$\\
2007 Oct 10.58079 &  0.42719 &  300$\times$4 &  2.3  & $R$    & Sampurnanand &  $20.00 \pm 0.20$\\
2007 Oct 11.04311 &  0.88950 & 60$\times$1     & 1.2     & $R$    & VLT+FORS2  & $19.78\pm0.04$ \\
2007 Oct 11.06735 &  0.91374 & 180$\times$1    & 1.3     & $R$    & VLT+FORS1  & $19.83\pm0.04$ \\
2007 Oct 11.19998 &  1.04637 & 30$\times$1 & 1.7      & $R$    & Keck\_I+LRIS & $20.05\pm0.04$\\
2007 Oct 11.20162 &  1.04801 & 30$\times$1 & 1.7      & $R$    & Keck\_I+LRIS & $20.01\pm0.05$\\
2007 Oct 11.20347 &  1.04986 & 45$\times$1 & 1.7      & $R$    & Keck\_I+LRIS & $20.03\pm0.04$\\
2007 Oct 11.20582 &  1.05221 & 45$\times$1 & 1.7      & $R$    & Keck\_I+LRIS & $20.04\pm0.04$\\
2007 Oct 12.02961 &  1.87600 & 60$\times$1     & 1.1     & $R$    & VLT+FORS2  & $21.34\pm0.04$ \\
2007 Oct 13.01453 &  2.86092 & 60$\times$5     & 1.1     & $R$    & VLT+FORS2  & $22.14\pm0.04$ \\
2007 Oct 14.00919 &  3.85558 & 60$\times$5     & 1.1     & $R$    & VLT+FORS2  & $22.87\pm0.05$ \\
2007 Oct 16.01841 &  5.86480 & 60$\times$15    & 1.2     & $R$    & VLT+FORS2  & $23.80\pm0.08$ \\
2007 Oct 16.20948 &  6.05586 &  3$\times$150   & 1.7     & $R$    & Keck\_I+LRIS & $24.03\pm0.05$\\
2007 Oct 10.82711 &  0.67350 & 0.1$\times$6000 & 3.2     & $I$    & NOT+PolCor & $18.81\pm0.05$ \\
2007 Oct 11.04457 &  0.89096 & 60$\times$1     & 1.2     & $I$    & VLT+FORS2  & $19.14\pm0.03$ \\
2007 Oct 12.03223 &  1.87862 & 60$\times$2     & 1.2     & $I$    & VLT+FORS2  & $20.65\pm0.03$ \\
2007 Oct 14.01544 &  3.86183 & 60$\times$5     & 1.1     & $I$    & VLT+FORS2  & $22.23\pm0.05$ \\
2007 Oct 16.01565 &  5.86204 & 60$\times$15    & 1.1     & $I$    & VLT+FORS2  & $23.06\pm0.10$ \\
2007 Oct 27.00014 & 16.84653 & 180$\times$3    & 1.2     & $I$    & VLT+FORS2  & $24.41\pm0.29$ \\
2007 Nov 11.01443 & 30.86082 & 180$\times$9    & 1.5     & $I$    & VLT+FORS2  & $24.44\pm0.25$ \\
2007 Oct 10.15742 &  0.00381 & 10$\times$5     & 2.3     & $J$    & REM+REMIR  & $14.72\pm0.18$ \\
2007 Oct 10.16498 &  0.01137 & 30$\times$5     & 2.5     & $J$    & REM+REMIR  & $15.08\pm0.13$ \\
2007 Oct 10.16737 &  0.01376 & 240                    &  2.5     & $J$  & GROND    & $15.16 \pm 0.11$ \\
2007 Oct 10.17856 &  0.02495 & 60$\times$5     & 3.0     & $J$    & REM+REMIR  & $15.64\pm0.13$ \\
2007 Oct 10.26426 &  0.11053 & 30$\times$9   &  2.0    & $J$    & GEMINI\_N+NIRI & $16.90\pm0.05$ \\
2007 Oct 11.15218 & 0.99856 & 1200                 &  2.2    & $J$   & GROND & $17.76\pm 0.10$ \\
2007 Oct 13.01086 &  2.85726 & 60$\times$15    & 1.1     & $J$    & NTT+SofI   & $19.90\pm0.08$ \\
2007 Oct 10.15519 &  0.00158 & 10$\times$5     & 2.3     & $H$    & REM+REMIR  & $14.43\pm0.15$ \\
2007 Oct 10.16038 &  0.00677 & 30$\times$5     & 2.4     & $H$    & REM+REMIR  & $14.02\pm0.10$ \\
2007 Oct 10.16737 &  0.01376 & 240                    &  2.5     & $H$  & GROND    & $14.38 \pm 0.05$ \\
2007 Oct 10.17049 &  0.01688 & 60$\times$5     & 2.8     & $H$    & REM+REMIR  & $14.52\pm0.10$ \\
2007 Oct 10.27297 &  0.11936 & 60$\times$12   &  2.1    & $H$  & GEMINI\_N+NIRI & $16.12\pm0.05$ \\
2007 Oct 10.82059 &  0.66698 & 60$\times$30    & 2.1     & $H$    & TNG+NICS   & $16.62\pm0.04$ \\
2007 Oct 11.15218 & 0.99856 & 1200                 &  2.2    & $H$   & GROND & $16.94\pm 0.05$ \\
2007 Oct 13.04876 &  2.89515 & 60$\times$20    & 1.2     & $H$    & NTT+SofI   & $19.01\pm0.07$ \\
2007 Oct 10.15632 &  0.00271 & 10$\times$5     & 2.3     & $K$    & REM+REMIR  & $13.42\pm0.20$ \\
2007 Oct 10.16270 &  0.00909 & 30$\times$5     & 2.4     & $K$    & REM+REMIR  & $13.39\pm0.12$ \\
2007 Oct 10.16737 &  0.01376 & 240                    &  2.5     & $K$  & GROND    & $13.66 \pm 0.07$ \\
2007 Oct 10.17455 &  0.02094 & 60$\times$5     & 2.8     & $K$    & REM+REMIR  & $14.04\pm0.14$ \\
2007 Oct 10.28911 &  0.13550 & 30$\times$6   &  2.3    & $K$     & GEMINI\_N+NIRI & $15.31\pm0.05$ \\
2007 Oct 11.15218 & 0.99856 & 1200                 &  2.2    & $K$   & GROND & $16.10\pm 0.05$ \\
2007 Oct 13.02786 &  2.87425 & 60$\times$25    & 1.1     & $K$    & NTT+SofI   & $18.20\pm0.06$ \\
2007 Oct 10.15523 &  0.00161 & 31              & 2.3     & clear  & TAROT      & $17.57\pm0.15$ \\
2007 Oct 10.15564 &  0.00203 & 29              & 2.3     & clear  & TAROT      & $17.20\pm0.10$ \\
2007 Oct 10.15793 &  0.00432 & 90              & 2.3     & clear  & TAROT      & $16.52\pm0.10$ \\
2007 Oct 10.15904 &  0.00543 & 90              & 2.3     & clear  & TAROT      & $16.39\pm0.10$ \\
2007 Oct 10.16023 &  0.00661 & 91              & 2.3     & $R$    & TAROT      & $16.60\pm0.10$ \\
2007 Oct 10.16138 &  0.00777 & 90              & 2.4     & clear  & TAROT      & $16.68\pm0.10$ \\
2007 Oct 10.16249 &  0.00888 & 89              & 2.4     & clear  & TAROT      & $16.82\pm0.10$ \\
2007 Oct 10.16367 &  0.01006 & 89              & 2.5     & $R$    & TAROT      & $16.97\pm0.10$ \\
2007 Oct 10.16484 &  0.01123 & 90              & 2.5     & clear  & TAROT      & $17.00\pm0.10$ \\
2007 Oct 10.16595 &  0.01234 & 91              & 2.5     & clear  & TAROT      & $17.10\pm0.10$ \\
2007 Oct 10.16714 &  0.01353 & 90              & 2.5     & $R$    & TAROT      & $17.03\pm0.10$ \\
2007 Oct 10.16829 &  0.01468 & 90              & 2.6     & clear  & TAROT      & $17.20\pm0.10$ \\
2007 Oct 10.16940 &  0.01579 & 90              & 2.6     & clear  & TAROT      & $17.16\pm0.10$ \\
2007 Oct 10.17059 &  0.01698 & 91              & 2.6     & $R$    & TAROT      & $17.27\pm0.12$ \\
2007 Oct 10.18639 &  0.03278 & 365             & 3.3     & clear  & TAROT      & $17.91\pm0.12$ \\
2007 Oct 10.19298 &  0.03936 & 365             & 3.6     & clear  & TAROT      & $17.98\pm0.12$ \\ 
\hline
\end{tabular}
\end{tiny}
\end{table*}

The {\it Swift} XRT data were reduced using the {\it xrtpipeline} task
(v.0.9.9), applying standard calibration and filtering criteria, i.e.,
we cut out temporal intervals in which the CCD temperature was above
$-47$\,$^\circ$C and removed hot and flickering pixels. An on-board
event threshold of $\sim 0.2$\,keV was applied to the central pixel;
this was proven to reduce most of the background due to the bright
Earth and/or CCD dark current. We selected XRT grades 0--12 for data
all in PC mode. A circular region of 20 pixel radius was selected for
source extraction. Physical ancillary response files were generated
using the task {\em xrtmkarf}. Optical/NIR and X-ray light curves are
shown in Fig.\,\ref{fig:lc_all}.

The \textit{Swift} BAT data were also analyzed. The 15--150\,keV
mask-weighted energy spectrum was extracted with the ftool {\it
batbinevt} using the BAT refined position \citep{Kri07}. All the
required corrections were applied: we accounted for the slewing
through the ftool {\it batupdatephakw} and produced the
detector-response matrix with {\it batdrmgen}. The spectrum was
corrected for systematics depending on energy with the ftool {\it
batphasyserr} and was finally grouped by imposing a 3$\sigma$
threshold on each grouped energy channel.

\subsection{Analysis}

The optical/NIR light curve is complex, with an initial rising phase,
a maximum, then a decay interrupted by a rather sharp rebrightening,
followed by a final steepening. We choose to fit the light curve with a
Beuermann function \citep[Eq.\,\ref{eq:beu},][]{Beu99} for the first
part down to the rebrightening:

\begin{equation}
F(t) = A/ {\left[ (t/t_{\rm b})^{\kappa\alpha_{\rm r}} + (t/t_{\rm b})^{\kappa\alpha_{\rm d}} \right]^{1/\kappa}},
\label{eq:beu}
\end{equation}
where $A$ is a normalisation constant, $\alpha_{\rm r(d)}$ is the
slope of the rise (decay) phase, and $\kappa$ is a smoothness
parameter. The time at which the curve reaches its maximum is $t_{\rm
max} = t_{\rm b} (-\alpha_{\rm r}/\alpha_{\rm
d})^{1/[\kappa(\alpha_{\rm d}-\alpha_{\rm r})]}$.

We model the rebrigthening at about 0.6\,d with a simple step
function. We also tried more complex models by using a pulse following
the functional form discussed in \citet{Nor96} and in
\citet{Gui07a}. The quality of data did not allow us to study the
transition in detail; we report the results requiring the minimum
number of free parameters. A general discussion is given in
Sec.\,\ref{sec:disc}.

The later-time evolution of the afterglow shows a final steepening,
and this was also modelled with a Beuermann function
(Eq.\,\ref{eq:beu}). Both for the onset and the final steepening the transition 
from the first to the second power-law behaviour was remarkably sharp and the
Beuermann's smoothness parameter had a high value ($\kappa \ge 10$,
frozen in the fits), making the model essentially consistent with what
would be obtained by using simple power-law segments. The temporal
decay shown by the {\it Swift}-XRT data was modelled with the same
functional form as the optical/NIR where quasi-simultaneous data were
available.

\begin{figure}
\includegraphics[width=\columnwidth]{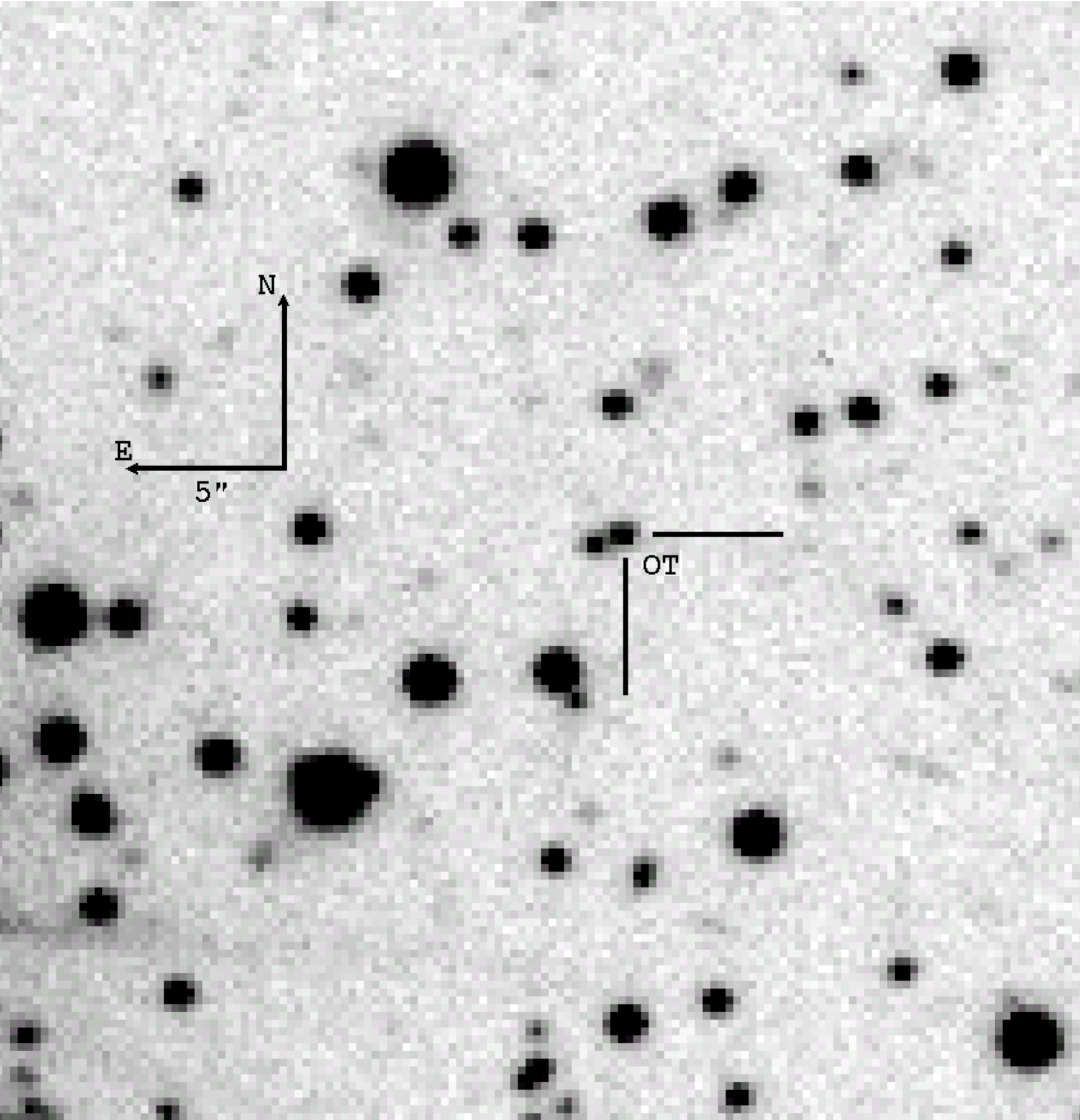}
\caption{The field of GRB\,071010A at about 3.8\,d after the GRB in
the $R$ band observed with the VLT equipped with FORS\,2. A nearby
source at $R=23.22\pm0.06$ at about one arcsec Eastward of the
afterglow is visible (see Sect.\,\ref{sec:dana}). The seeing of the
image is $\sim 0.7$\,arcsec.}
\label{fig:field}
\end{figure}

The data were fitted both in the spectral and temporal domains. The
optical/NIR afterglow spectrum was modelled with a simple power
law. Galactic dust absorption was removed and rest-frame absorption
was modelled with the absorption curves of the Milky Way (MW), the
Large and Small Magellanic Clouds (LMC, SMC) \citep{Pei92}, and for a
starburst galaxy (SB) \citep{Cal00}. For the $I$ band we also added a
constant component, possibly the host galaxy or a supernova (SN), as
late-time data (later than about 10\,d) show a clear
flattening. However, the presence of a red point-like object $\sim1''$
east of the afterglow, with $R = 23.22 \pm 0.06$ mag and $I = 22.31
\pm 0.07$ mag, makes late-time photometry less reliable (see
Fig.\,\ref{fig:field}). This source could also contribute to the NIR
flux at about 2.8\,d, when the observing conditions did not allow a
reliable separation. A simple extrapolation to the NIR bands for $R-K
\approx 1$ mag would suggest that it could be comparable in brightness
to the afterglow. We therefore did not consider these data in the
fits.

The model gives an acceptable fit ($\chi^2_{\rm dof} = 114.1/90 \approx
1.27$), particularly if one considers the number of telescopes
contributing the dataset under different observing conditions,
filters, etc.  A summary of the best-fitting parameters is reported in
Table\,\ref{tab:fitpar}.

\begin{table}
\caption{Best-fitting parameters.}
\centering
\begin{tabular}{cc}
\hline
\hline
Parameter & Interval \\
\hline
$\alpha_{\rm r}$        & $-0.88^{+0.43}_{-0.62}$  \\
$t_{\rm max}$           & $420_{-85}^{+124}$\,s   \\
$\alpha_{\rm d_1}$ & $0.71_{-0.04}^{+0.03}$  \\
$t_{\rm break}$         & $0.96_{-0.09}^{+0.09}$\,d   \\
$\alpha_{\rm d_2}$ & $2.07_{-0.07}^{+0.08}$  \\
$f_{\rm inj}/f$ & $\sim 1.9$           \\
$\beta_{\rm opt/NIR}$   & $0.76_{-0.26}^{+0.23}$\\
$E_{B-V}$   & $0.21_{-0.05}^{+0.05}$ \\
$\beta_{\rm X}$  & $1.46_{-0.18}^{+0.22}$  \\
$N_{\rm H}$        & $1.74(_{-0.37}^{+0.57}) \times 10^{22}$\,cm$^{-2}$ \\
\hline
\end{tabular}
\label{tab:fitpar}
\end{table}

The early-time observations by TAROT, REM, and GROND allowed us to
identify a rising phase, a maximum, and then a regular decay
phase. The fit yields $t_{\rm max} = 420^{+124}_{-85}$\,s ($\sim
7$\,min), $\alpha_{\rm r} = -0.88_{-0.62}^{+0.43}$, and $\alpha_{\rm
d_1} = 0.71^{+0.03}_{-0.04}$, for the time of the maximum and the
rising and decaying power-law indices, respectively, in substantial
agreement with the findings reported by \citet{Klo07b}. The data 
do not suggest the presence of any spectral evolution.

\begin{figure*}
\begin{tabular}{cc}
\includegraphics[width=\columnwidth]{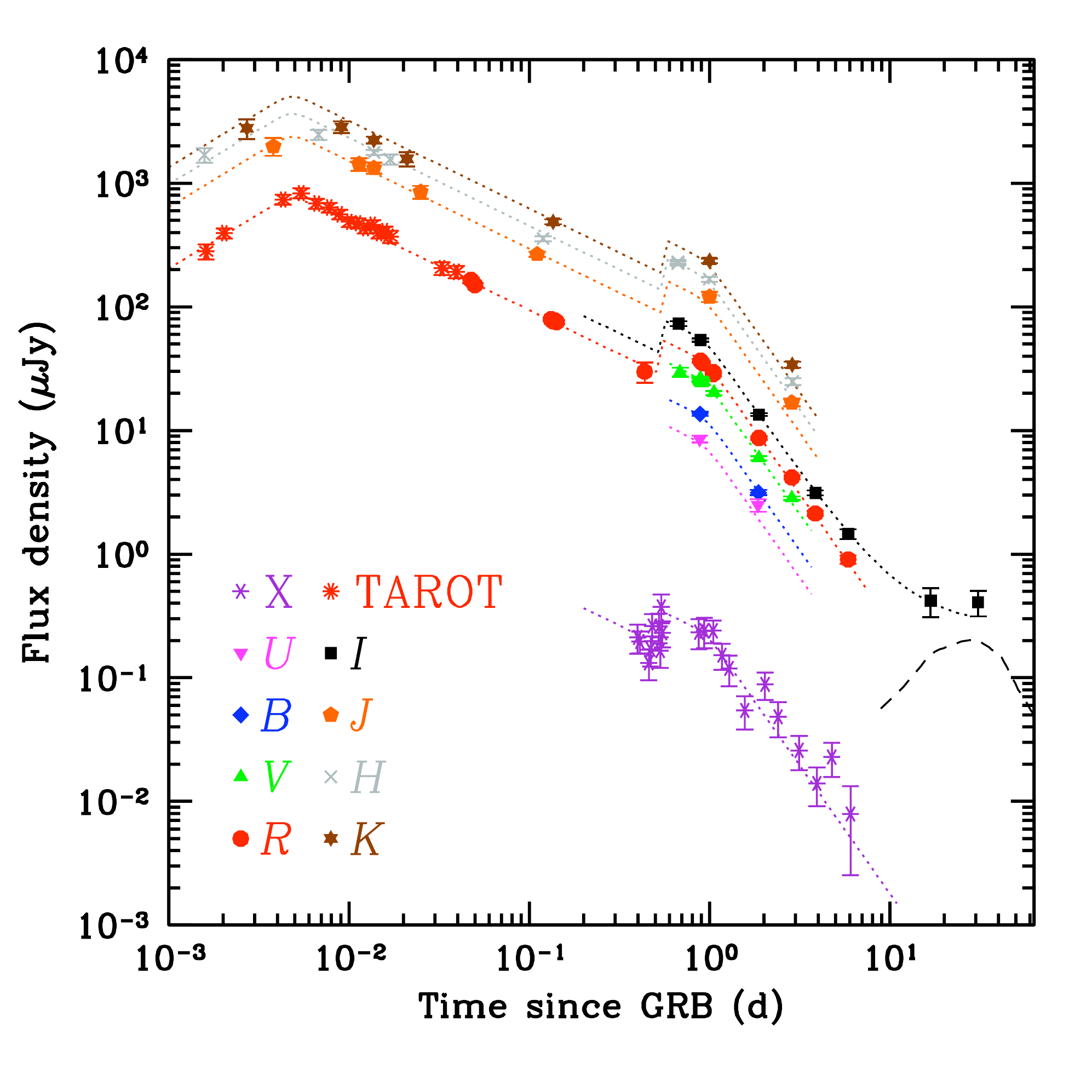} &
\includegraphics[width=\columnwidth]{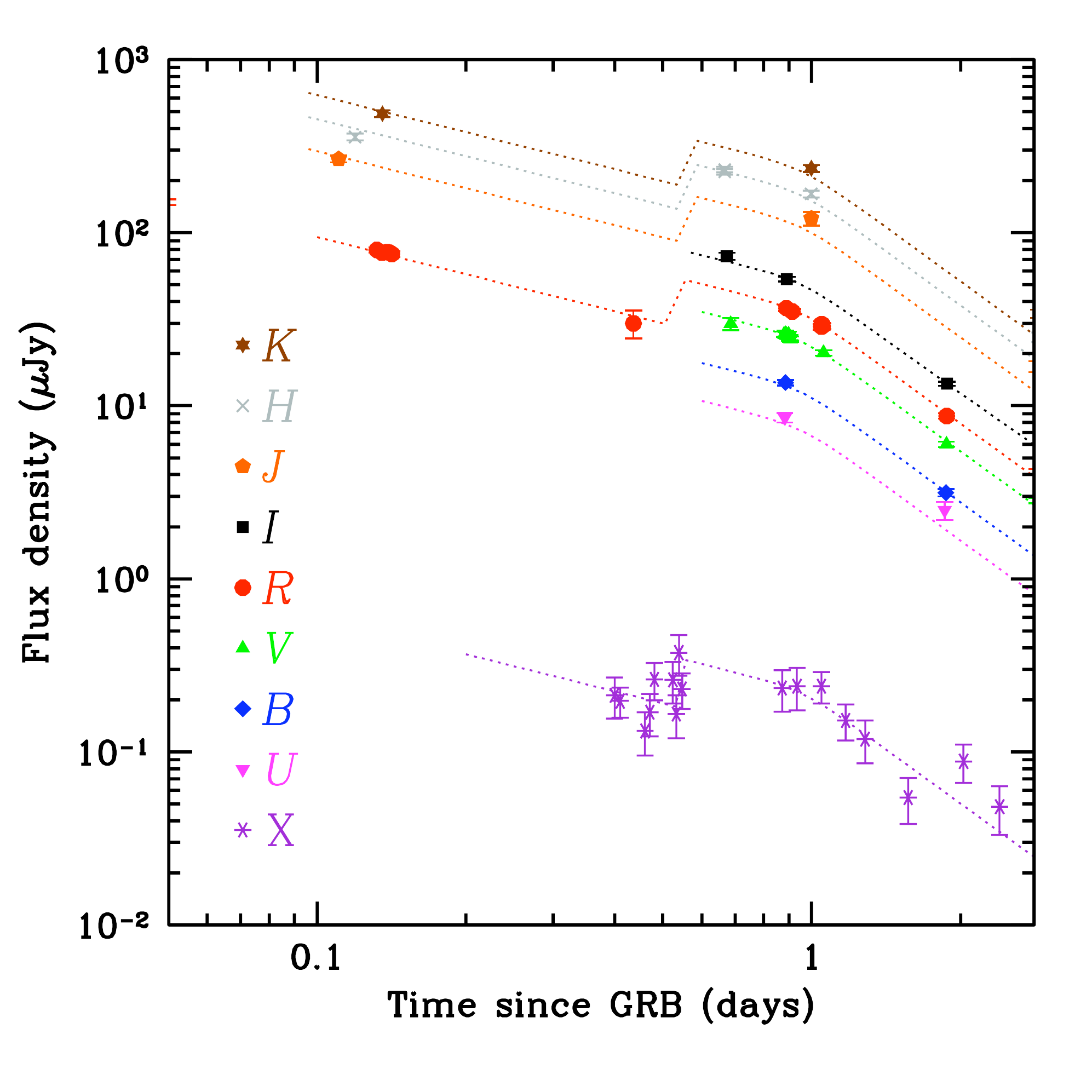}
\end{tabular}
\caption{The optical, NIR, and X-ray light curves of the afterglow of
GRB\,071010A. The dashed lines show the fit to the temporal decay of
the early and late-time afterglow. Fit parameters are reported in
Sect.\,\ref{sec:dana}. On the left we show the whole light curve
fitted with the adopted model. On the right the epoch around $1$\,d is
magnified. At late times, the long-dashed line shows the possible 
contribution to the $I$ band by a supernova component using SN\,1998bw 
\citep{Gal98} as a template, at the redshift of GRB\,071010A. The 
effect of the absorption in the optical is included.}
\label{fig:lc_all}
\end{figure*}

A rebrightening peaking at about $0.6$\,d dominates the afterglow at
intermediate times. The required flux ratio increase ($\sim 1.9$) is
remarkably high and, within the uncertainties, the same in the
optical/NIR and X-rays. The later evolution of the afterglow requires
a break, consistent with being fully achromatic in the optical/NIR and
X-ray bands, at $t_{\rm break} = 0.96_{-0.09}^{+0.09}$\,d, with a
post-break decay index of $\alpha_{\rm d_2} = 2.07_{-0.07}^{+0.08}$,
followed by a final flattening likely due to the contribution from the
GRB host galaxy. The best-fitting model is shown superposed on the
data in Fig.\,\ref{fig:lc_all}.

The entire afterglow spectral evolution is consistent with being
achromatic. In the optical/NIR the broad-band spectral energy
distribution index is $\beta = 0.76_{-0.26}^{+0.23}$, with a sizeable
rest-frame extinction $E_{B-V} = 0.21_{-0.05}^{+0.05}$ mag after
having removed the Milky Way absorption along the line of sight. Among
the various extinction curves, only that of the SMC gave satisfactory
results. The redshift of GRB\,071010A would move the 2175~\AA\ bump
\citep[see ][for a discussion about observations of this feature in
GRB afterglow spectra]{Fyn07}, typical of Milky Way absorption
\citep{Pei92}, into the $B$ band, therefore easily detectable, if
present, in our data sets.

The X-ray spectrum was modelled assuming a simple power law with
neutral absorption in the Milky Way and in the host galaxy.  Analysis
of the hardness ratio showed that there are no significant spectral
variations during the X-ray observations. The spectral analysis of the
X-ray data alone ($\chi^2_{\rm dof} = 14.4/19 \approx 0.76$) provided
satisfactory results. Fixing the Galactic absorption at $6.5 \times
10^{20}$\,cm$^{-2}$ \citep{DiLo90} the rest-frame absorption and the
spectral index turned out to be $1.74(_{-0.37}^{+0.57}) \times
10^{22}$\,cm$^{-2}$ and $\beta = 1.46_{-0.18}^{+0.22}$, respectively. 
The optical/NIR and X-ray spectral
indices are consistent within the uncertainties with $\beta_{\rm X}
\approx \beta_{\rm opt} + 0.5$, i.e. with the cooling-break between
the two bands. The broad-band SED is shown in Fig.\,\ref{fig:sed}.

The prompt time-averaged spectrum of GRB\,071010A in the
\textit{Swift} BAT 15--150\,keV energy band is consistent with a single
power law with photon index significantly higher than 2 \citep{Kri07},
suggesting that the peak energy, $E_{\rm peak}$, of this event is
close to, or below, 20--30\,keV. In order to better investigate this
issue, we reduced and analyzed the BAT time-averaged spectrum. The fit
with a simple power law yelds a photon index of 3.1$\pm$0.7,
confirming that GRB\,071010A is a soft event, an X-ray flash
(XRF). By fitting the spectrum with a Band function \citep{Band93}
with $\alpha$ fixed at $-$1, we derive an upper limit to the peak
energy $E_{\rm peak}$ of $\sim$35\,keV, which corresponds to
$\sim$70\,keV in the GRB cosmological rest frame. Then, by varying
$E_{\rm peak}$ from its upper limit to $0$ and adopting the
method of \citet{Ghi04} and \citet{Ama06}, we estimated an
isotropic-equivalent radiated energy in the 1--10000\,keV cosmological
rest-frame energy band, $E_{\rm iso}$, of $(3.6 \pm 2.9) \times
10^{51}$\,erg.

\section{Discussion}
\label{sec:disc}

\subsection{The onset}

The temporal resolution of the observations does not allow us to check
in detail the early-time density profile of the circumburst medium or
possible chromaticity in the evolution. The modest duration of the
prompt event and the late onset time argues against a scenario with
frequent multiple peaks as for GRB\,070311 \citep{Gui07a}. A smooth
rising phase can also be produced by a decreasing extinction in the
case of a radially decreasing circumburst density
\citep{Ryk04}. However, the detection of the afterglow in the $U$ band
(Table\,\ref{tab:photo}) eliminates this last possibility due to the
high absorption required.  An alternative giving a chromatic maximum
could be the passage of the typical synchrotron frequency in the
optical/NIR bands. The time spread for the maximum would depend on the
width of the observed frequency band, $t_1/t_2 \propto (\nu_1 /
\nu_2)^{-2/3}$, with the higher frequencies peaking first. The passage
would also produce a spectral change from a positive ($1/3$) to
negative $[-(p-1)/2]$ spectral index in the slow-cooling case
\citep{Sar98,ChLi99}. Again, this is not supported by the data.

The maximum during the early afterglow evolution can then be
interpreted as the afterglow onset as for the case of GRB\,060418 and
GRB\,060607A \citep{Mol07,JiFa07}. Following the formalism reported in
Eqs.\,1 and 2 of \citet{Mol07} in the thin-shell case since the prompt duration 
was much shorter than the peak of the afterglow emssison, we can derive,
for GRB\,071010A, $\Gamma_0 \approx 150$ for a uniform interstellar
medium (ISM) and $\Gamma_0 \approx 40$ in the case of a wind-shaped
density profile. These are lower than the values inferred for
GRB\,060418 and GRB\,060607A due to the late occurrence of the
maximum, but still within the theoretical expectations for the
external-shock scenario \citep{SaPi99,LiSa01,Zha06}. 

The observed maximum cannot be related to reverse shock emission. The reverse
shock should be in general a short-lasting phenomenon, and in the thin-shell case 
it should peak slightly before the deceleration time, roughly just before the afterglow 
onset, and dominate it in case the typical synchrotron emission is close enough to the 
optical band \citep{KoZha07,JiFa07}. The rapid decay \citep[$f \sim t^{-2.1}$, ][]{SaPi99,KoZha07}
predicted for the reverse shock emission is inconsistent with the much milder decay observed 
after the optical flux peak (Table\,\ref{tab:fitpar}).

\subsection{The rebrightening}

The occurrence of rebrightenings or flares during the temporal evolution
of GRB afterglows is not uncommon. The long time coverage usually
provided by the \textit{Swift} XRT affords the study of a large sample
of flares from the spectral and energetic \citep{Fal07}, or temporal and
morphological \citep{Chi07} points of view. In the optical/NIR the
occurrence of large flares seems to be less common, while in a
remarkable fraction of cases a rebrightening due to discrete energy
injection was included in the modelling \citep{Joh06}.

Our dataset does not permit a detailed analysis of the temporal and
spectral evolution of the rebrightening around 0.6\,d. The simplest
approach is to model the transition as due to the discrete injection
of a sizeable amount of energy in the fireball interacting with the
external medium by means of refreshed shocks. The flux density
produced by an external shock scales with the energy content of the
fireball apart from minor differences between the optical/NIR and
X-ray bands \citep{PaKu00} which are not detectable in our data.
Therefore, to account for the observed $ f_{\rm inj}/f \approx 1.9$,
an amount of energy comparable to the energy content of the fireball
needs to be supplied to the system. Remarkably, a discrete episode of
energy injection is not supposed to modify the spectrum of the
synchrotron radiation emitted at the shock front if the cooling
frequency does not enter or cross the observing bands \citep{PaKu00},
in agreement with the achromatic evolution of the afterglow of
GRB\,071010A. A discrete energy injection, comparable to the energy
content of the fireball, was also suggested by \citet{deUg07} for
GRB\,050408. Also in the case of the XRF\,050824 a rebrightening was
observed at rather late time and a large delayed energy injection was
invoked \citep{Sol07}.  In principle, a rebrightening could be
induced by a density variation in the circumburst medium as proposed
by \citet{Laz02} for the GRB\,021004 \citep[but see also][]{NaGr07}.
However, this explanation cannot be applied to our case, since the
X-ray range is above the cooling frequency and therefore insensitive
to density variations \citep[see, e.g.,][]{Sar98}. A transition from wind 
to homogenous medium, due to the crossing of the wind reverse shock 
could also generate a rebrightening, as discussed in \citet{PeWi06}.
Other explanations are also possible, such as a double jet \citep{Berg03} or a patchy shell
\citep{ReMe98,Mes98}. A structured jet with a bright narrow core and
fainter wings can also produce a rebrightening close to the jet-break
time \citep{Sal03}, though in such a case it would only be due to
perspective rather than to real energy injection, possibly driving to
a sharp transition if seen from a suitable angle.

Finally, although not required by the data, the temporal profile of
the rebrightening is also consistent with that of a fast rise and
exponential decay pulse simply superposed on the underlying
afterglow. The relatively long duration of the pulse, as compared to
the time of occurrence, $\Delta t/t \approx 1$, is compatible with an
origin of the emission at the external-shock radius \citep[see
discussion in][]{Gue07,Gui07a}.

\subsection{Late-time afterglow, light-curve break, and SED analysis}

Following the rebrightening at about 0.6\,d, the afterglow evolution
shows a sharp break at about 1\,d consistent with being achromatic in
the X-ray and optical/NIR bands (Fig.\,\ref{fig:lc_all}). Following
\citet{Lian07} only seven GRB afterglows have been found so far with
such an achromatic transition. Moreover, the general temporal
behaviour of the optical/NIR and X-ray light curves seems to be the
same, supporting at least qualitatively the possibility that we are
observing emission from a single outflow with a transition due to a
beamed geometry.

\begin{figure}
\includegraphics[width=\columnwidth]{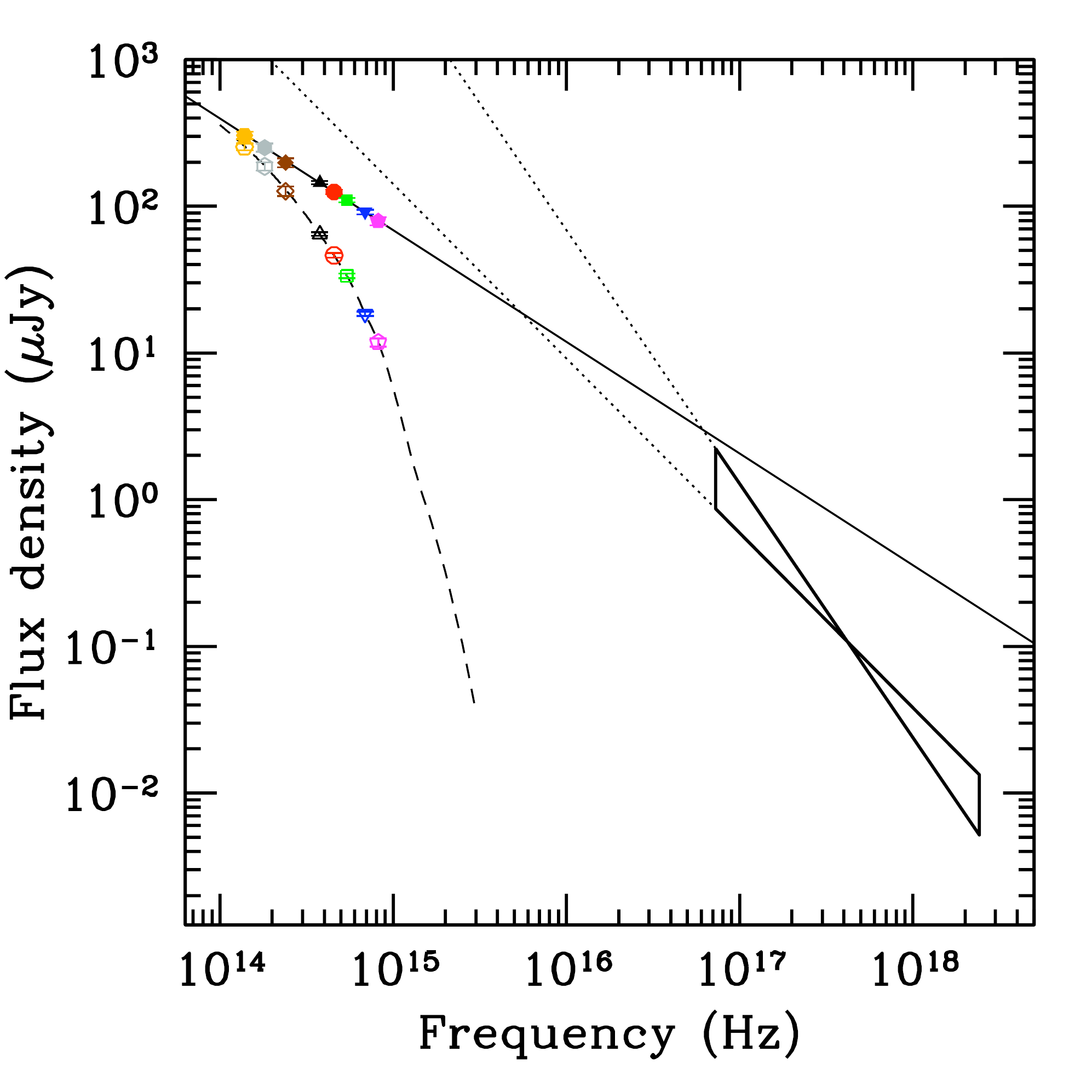}
\caption{SED of the afterglow of GRB\,071010A about 0.8\,d after the
burst. The dashed line is a fit to the observed data in the
optical/NIR. The solid line refers to the intrinsic SED once the
effect of dust absorption in the Milky Way and in the GRB host galaxy
is removed. 
The X-ray and optical/NIR broad-band SED is consistent with the predictions
 of the synchrotron external-shock scenario with a cooling break between the optical/NIR
 and the X-ray and $\beta_{\rm X} = \beta_{\rm opt} + 1/2$.}
\label{fig:sed}
\end{figure}

The standard external-shock model predicts specific relations between
temporal and spectral parameters. However, it is a well-established
observational fact that GRB afterglows show more complex behaviours
than expected before the launch of \textit{Swift}. Many mechanisms
have been proposed to interpret the new features discovered with
\textit{Swift}: delayed energy injection, late central engine
activity, refreshed shocks, high-latitude emission,
etc. \citep[e.g.,][]{ZhMe04,Zha06}. Most of these ingredients affect
the light curve shape in a number of ways. However, since the
fundamental emission process in the soft X-rays and in the optical/NIR
is still synchrotron, the spectra should follow the general
predictions of the external-shock scenario. We therefore begin the
analysis starting from the broad-band SED from the optical/NIR to the
X-rays. The evolution of the afterglow, including the early phases, is
consistent with being fully achromatic. We derive the SED at about
0.8\,d in order to minimise the need to extrapolate the observations
adopting a temporal behaviour (Fig.\,\ref{fig:sed}).

The amount of rest-frame neutral absorption required by the X-ray data
($N_{\rm H} \approx 1.7\times10^{22}$\,cm$^{-2}$) is within the range
observed for a set of well-studied \textit{Swift} GRB X-ray afterglows
\citep{Cam06}. In the optical/NIR the observed SED curvature clearly
indicates the need for a substantial absorption, as derived in
Sect.\,\ref{sec:dana}, with $E_{\rm B-V} \approx 0.21$ mag. Such an
amount of intrinsic absorption in the optical/NIR is uncommon among
GRBs with sufficient data quality to allow a meaningful study
\citep[see ][for a comprehensive
analysis]{Stra05,Tag06,Sta07,Kann07}. As for most of the GRB optical
afterglows, a SMC-like extinction curve seems to provide a better fit
to the data \citep{Kann06,Kann07}, although the gas-to-dust ratio
$N_{\rm H} / E_{B-V}$ is higher than in the SMC, again in
agreement with the findings of \citet{Sta07}.

The spectral indices in the optical/NIR and X-ray bands allows us to constrain the
electron energy distribution index $p$ ($N \propto \gamma^{-p}$). With
the cooling frequency in between the optical/NIR and the X-rays
(probably just below the X-ray range, as shown in
Fig.\,\ref{fig:sed}), we have $\beta_{\rm opt} = (p-1)/2$ and $\beta_{\rm X} = p/2$
and therefore $p_{\rm opt} = 2.52^{+0.46}_{-0.52}$ and $p_{\rm X} = 2.92^{+0.44}_{-0.36}$, independent of the circumburst matter-density profile in the slow-cooling regime 
\citep[][and references therein]{ZhMe04}. These values for $p$, in particular from the 
X-rays, lit in the high tail of the distribution derived by \citet{Tag06} or \citet{She06}, although still
consistent with it. The value derived from the optical is also consistent with the typical results
($p \approx 2.2$) of numerical simulations of relativistic shocks
\citep{Acht01,Vie03}. Knowing the electron distribution index we can
predict the decay rate in the optical/NIR before the occurrence of a
jet break, considering also the constraints from the X-rays: 
$\alpha_{\rm pre,opt} \approx 1.2-1.5$ (0.75-1.5 from the optical/NIR only) 
in the case of a constant-density radial profile of the circumburst matter, or
$\alpha_{\rm pre,opt} \approx 1.7-2.0$ (1.25-2.0 from the optical/NIR only) 
in the case of an environment shaped by a wind from a massive star. 
For the X-rays (frequencies above the cooling frequency) we have 
$\alpha_{\rm pre,X} \approx 1.4-1.75$ (1.4-2.0 from the X-ray only)
 independent of the density profile (and unconstrained by our data).

It is clear from the fit results, reported in Sect.\,\ref{sec:dana}
and shown in Fig.\,\ref{fig:lc_all}, that interpreting the break at
$\sim 1$\,d as a jet break, the optical/NIR decay before the
transition would be in essential agreement with the expectations for a
constant-density medium only for the $p$ value derived from the optical.  
Otherwise, a continuous additional injection of energy as required by the modeling 
of numerous \textit{Swift} X-ray afterglows \citep{Pan06} should be considered. 
After the jet break, 
the decay index in the slow-cooling regime must be the same at all frequencies 
(higher than the typical synchrotron frequency, as expected $\sim 1$\,d after the
high-energy event), and independent of the circumburst matter-density
profile. For $p > 2$, as in our case, it is simply $\alpha_{\rm post}
= p$. Again the observed decay indices are consistent with the
expectations only with the value derived from the optical/NIR.

In principle, the post-break decay indices would also be consistent
with a wind model before the occurrence of a jet break with $p \approx
3.1$. This $p$ value would be better supported by the X-ray spectral data, 
although the achromaticity of the transition at $\sim 1$\,d is not consistent
with this scenario. The distribution of $p$ values, although likely 
inconsistent with a unique $p \approx 2.2$ \citep{Tag06,She06}, supports $p < 3$ 
for essentially all the afterglows once the analysis is based on the spectral parameters
rather than on the temporal behaviour.

Summing up, the hypothesis that the sharp achromatic transition at
about $1$\,d be a real jet break appears consistent with the data if we 
assume that continuous energy injection is affecting the temporal decays. 
This would be one of the very few instances of a jet break observed both in
the optical/NIR and the X-rays. The consistency of the GRB\,071010A
parameters with the Ghirlanda relation (see below) lends some indirect
support to the jet-break interpretation.

After about ten days the afterglow light curve flattens considerably,
plausibly as a result of the emergence of the host-galaxy light.  Our
data quality does not permit to draw any conclusion about the possible
presence of a SN component affecting the light curve and mimic the
host-galaxy contribution (see Fig.\,\ref{fig:lc_all}). A SN component
comparable to SN\,1998bw \citep{Gal98} at the redshift of GRB\,071010A
and including the effect of the fitted absorption in the optical would
be in the $I$ band about 0.8\,mag fainter than our last measured
point.

\subsection{Amati and Ghirlanda relations}

It is well known that the position of a GRB in the radiated
energy vs. spectral peak photon energy plane can provide useful clues on
the nature, physical origin, and geometry of its emission. Indeed,
while all ``normal" long GRBs and XRFs show a clear correlation
between these two quantities, short GRBs and peculiar sub-energetic
GRBs (980425 and, possibly, 031203) show a different behavior
\citep[e.g.,][]{Ama06}. In addition, collimated GRBs seen off-axis
are expected to deviate from these spectrum-energy correlations.

The $E_{\rm peak}$ and $E_{\rm iso}$ values derived from the analysis
of the time-averaged spectrum (see Sect.\,\ref{sec:dana}) are fully
consistent with the $E_{\rm p,i}-E_{\rm iso}$ (Amati) relation
\citep{Ama02,Ama06}, showing that the prompt emission of GRB\,071010A
is not peculiar and indicating that, if the emission is collimated,
the off-axis angle is small. In addition, with a break interpreted as
a jet break at about 1\,d and by assuming, following \citet{Ghi04}, a constant circumburst
matter-density profile ($n=3$\,cm$^{-3}$) and a fireball kinetic to
radiated energy conversion efficiency of 0.2, we derive a jet opening
angle of $8.2 \pm 1.1$ deg and a collimation-corrected energy of
($3 \pm 2) \times 10^{49}$\,erg. Thus, GRB\,071010A is also fully
consistent with the $E_{\rm p,i}-E_\gamma$ (Ghirlanda) relation
\citep{Ghi04,Nav06}. This evidence further supports that the possibly
achromatic break observed in the afterglow light curve is due to
collimated emission.

Finally, if we fit the time-averaged spectrum of GRB\,071010A with a
cut-off power law with index fixed at $-1$, we can constrain the
cosmological rest-frame peak energy to 32$_{-21}^{+46}$\,keV. This
value, combined with the values of $E_{\rm iso}$ and $E_\gamma$
derived above, make the representative point for GRB\,071010A
consistent with the best-fit power law of both the $E_{\rm p,i}-E_{\rm
iso}$ and $E_{\rm p,i}-E_\gamma$ relations, thus reinforcing the above
considerations.

\subsection{Polarisation}

Almost coincident with the achromatic break, we obtained polarimetric
observations, providing a rather robust upper limit (0.9\% at 95\%, 1.3\% at 3\,$\sigma$) 
suggesting null polarisation. Polarimetric measurements of GRB optical afterglows
have been performed for several events \citep{Cov04}. In general, the
polarisation level was moderate (2--3\%), although in a few cases
evidence for variations related to the afterglow evolution was found
\citep{Laz03,Ber03,Rol03,Grei03,Goro04,Laz04}. The simple detection of
variable polarization, intrinsically related to the afterglow
emission, is still one of the most important observational tests in
favour of the external relativistic shock model with physical beaming,
where some degree of polarisation is naturally expected
\citep{Mal05,Cov05,Laz06}.

Almost all of the observations have been performed with the goal of
disentangling the geometry of the outflow using time-resolved
polarimetry. After an initial period of enthusiasm, the required
intense and delicate observational efforts and the high degree of
complexity of the afterglow light curves provided by \textit{Swift}
cast some doubts on the actual possibility of deriving meaningful
constraints from simple models \citep{GhLa99,Sar99}. These proved to
be applicable only to smooth, regular, afterglow light
curves. Nevertheless, \citet{Ros04} showed that the polarisation
degree and position-angle evolution is strongly dependent on the
assumed jet structure. At least in a couple of cases a homogeneous jet
geometry could be formally ruled out: i.e. for GRB\,020813
\citep{Laz04} and for GRB\,030328 \citep{Mai06}.

The diagnostic power of time-resolved polarimetric observations relies
on the strong coupling between the assumed jet geometry and
polarization signature. For homogeneous jets there should be two
polarization maxima, before and after the jet-break time, with a
rotation of the position angle by 90$^\circ$. For structured jets,
(i.e., jets with a more energetic core and fainter wings), the maximum
of polarization should almost be coincident with the jet-break time,
and the polarization angle constant. Therefore, a polarimetric
measurement almost coincident with a jet break, as in the case for
GRB\,071010A, is of remarkable value. The very low upper limit that we
have obtained (Sect.\,\ref{sec:dana}) argues against a structured
jet. However, the lack of time-resolved polarimetry through the
evolution of the afterglow prevents us from drawing firm
conclusions. The polarised fraction for the radiation emitted by an
afterglow depends basically on the degree of asymmetry of the system,
which in turn depends on the line of sight of the observer with
respect to the jet symmetry axis. With a single observation the
possibility of a very small offset angle cannot be ruled out. Setting
aside the jet-break hypothesis, it is still true that, whichever
physical process is responsible for the (achromatic) steepening
observed in the afterglow at about 1\,d, it generated radiation with
only a very low level of polarization. The same is true if we model
the rebrightening at about 0.6\,d with a fast rise exponential
decline pulse dominating the afterglow
luminosity for the pulse duration.

\section{Conclusion}
\label{sec:end}

Our campaign on GRB\,071010A is an example of extensive monitoring of
a GRB afterglow in the optical/NIR. This allowed us to detect an
initial rising phase which can be interpreted as the afterglow
onset. The peak time is rather late, about 7\,min after the GRB. This
would directly imply a relatively small initial Lorentz factor
although still within the fireball model expectations.  Interesting
features are the rebrightening at $\sim 0.6$\,d followed by an
achromatic break $\sim 1$\,d after the burst. The steepening is
observed both in the optical/NIR and in the X-rays bands, a feature
seldom observed even in the \textit{Swift} era.  

The spectral and temporal relations before and after the break are roughly consistent
with the predictions of the afterglow synchrotron external
shock model with a cooling-break in between the optical/NIR and X-ray bands and
assuming also continuous energy injection. The rebrightening episode can be interpreted as
being due to the occurrence of a refreshed shock injecting energy in
the system comparably to the original energy content of the fireball. Other 
explanations are still possible, relating the rebrightening to the density profile
of the circumburst medium or to structured jets. The optical/NIR spectral energy distribution
requires absorption in the host-galaxy satisfactorily modeled 
with the SMC extinction curve.

GRB\,071010A is consistent with the Amati relation and, if the
achromatic steepening is interpreted as a jet break, it is consistent
with the Ghirlanda relation too. Independent of the emission process,
the entire afterglow evolution is consistent with being independent of
color from about 2\,min up to a few days. The SED is consistent with
synchrotron emission and requires rest-frame absorption both in the
X-rays and in the optical/NIR. 

We also performed a polarimetric observation almost in coincidence
with the occurrence of the achromatic break. A tight (0.9\%) upper 
limit was derived. In the absence of time-resolved polarimetry, we
cannot rule out that this afterglow was characterized by a low level
of polarization throughout its evolution.

Finally, the rich dataset of this burst and the simultaneous
availability of multi-wavelength data is a positive example of a new
approach we are trying to apply to GRBs localized by \textit{Swift},
favouring more complete coverage of a few well-studied GRBs. In
addition, the observation of an initial rising phase, a rebrightening
episode, and an achromatic break, as well as the availability of a
well-sampled broad-band energy distribution, could offer a valuable
workbench for alternative interpretative scenarios such as the
``cannonball" \citep{DeR07,Dad07}, the ``fireshell" \citep{Bia07}, and
the ``quark nova'' \citep{Staf07} models in order to reach a better
understanding of GRB afterglow phenomenology.

\section*{Acknowledgments}

SC thanks G. Ghirlanda, G. Ghisellini, D. Malesani and F. Tavecchio for useful
discussions. AVF's group at UC Berkeley is funded by NSF grant AST--0607485 and 
NASA/{\it Swift} grant NNG06GI86G. BG acknowledges a post-doctoral grant funded by 
the Centre National d'Etudes Spatiales. We also thank the anonymous referee for her/his 
useful comments and suggestions.

\label{lastpage}

\end{document}